\documentstyle[12pt,draft]{article}
\parskip =\medskipamount
\newfont{\ensmathquatorze}{msbm10 scaled 1400}
\newfont{\ensmathonze}{msbm10 scaled 1100}
\newfont{\ensmathdix}{msbm10}
\newfont{\ensmathneuf}{msbm10 scaled 833}
\newfont{\ensmathhuit}{msbm10 scaled 694}
\newfam\ensmathfam                        
\textfont\ensmathfam=\ensmathonze        
\scriptfont\ensmathfam=\ensmathdix       
\scriptscriptfont\ensmathfam=\ensmathhuit

\def\be{\begin{equation}}
\def\ee{\end{equation}}
\def\bea{\begin{eqnarray}}
\def\eea{\end{eqnarray}}

\newcommand\li{\lambda_i}
\newcommand\lj{\lambda_j}
\newcommand\vli{\vert\lambda_i\vert}
\newcommand\vlj{\vert\lambda_j\vert}
\newcommand\tdz{^t\partial_z}
\newcommand\dzi{\partial_{z_i}}
\newcommand\dzib{\partial_{\bar z_i}}
\newcommand\dz{\partial_z}
\newcommand\dzb{\partial_{\bar z}}
\newcommand\tdzb{^t\partial_{\bar z}}
\newcommand\ti{\theta_i}
\newcommand\tj{\theta_j}
\newcommand\bb{{\bf B}}

\newcommand\gM{{\bf M}}
\newcommand\gR{{\bf R}}

\newcommand\ddz{{\cal D}z}
\newcommand\ddzb{{\cal D}{\bar z}}
\newcommand\oi{\omega_i}
\newcommand\oj{\omega_j}

\newcommand\bp{{\bf B}'_{11}}
\newcommand\bpp{{\bf B}''_{11}}

\newcommand\zo{z^{(0)}}
\newcommand\zob{\overline{ z}^{(0)}}
\newcommand\dtm{\int_0^t {\rm d}t_m \int_0^{t_m}{\rm d}t_{m-1}
\ldots \int_0^{t_2} {\rm d}t_1 }
\newcommand\Lp{\left(} 
\newcommand\Rp{\right)} 
\newcommand\La{\left\langle} 
\newcommand\Ra{\right\rangle}
\newcommand\Lv{\left\vert}
\newcommand\Rv{\right\vert}
\newcommand\dd{{\rm d}}
\newcommand\AAA{{\cal A}} 

\begin{document}


\title{  \bf  Windings of the 2D free Rouse chain}

\author{ Olivier B\'enichou$^{\dagger ,\star }$ and Jean Desbois$^{\dagger }$ }

\maketitle	

{\small
\noindent$^\dagger $ 
Laboratoire de Physique Th\'eorique et Mod\`eles Statistiques.
Universit\'e Paris-Sud, B\^at. 100, F-91405 Orsay Cedex, France.
}

{\small
\noindent$^\star$ 
Laboratoire de Physique Th\'eorique des Liquides, UPMC, 4 place Jussieu
 75252 Paris Cedex 05 
}
\vskip1.3cm

\begin{abstract}

We study long time dynamical properties of a chain of harmonically bound
Brownian particles. This chain is allowed to wander everywhere in the plane.
We show that the scaling variables for the occupation times $T_j$, areas
$A_j$ and winding angles $\theta_j$ ($j=1,\ldots ,n$ labels the particles) 
take the same general form as in the
usual Brownian motion. We also compute the asymptotic joint laws
$P(\{T_j\}),\;P(\{A_j\}),\;P(\{\theta_j\})$ and discuss the correlations
occuring in those distributions.

 \end{abstract}

\vskip1.3cm

\section{Introduction}

In order to study the dynamics of dilute polymer solutions, P.E. Rouse
proposed in 1953 his famous model of harmonically bound Brownian particles
(Rouse chain) \cite{1}. Since that time, this model has become very popular
in the field of polymer science. It appears that, despite its drawbacks and
limitations (in particular, absence of excluded volume and hydrodynamic
interactions), it is conceptually important and useful to study the dynamics
of polymers in melts [2,3]. In this paper, we will consider the free
planar motion of such a chain of $n$ particles (monomers) and especially
address its long time ($t\to\infty$) properties from the Brownian motion
viewpoint.

A configuration of this chain being represented by a complex $n$-vector $z$
(the components $ z_i,$ $i=1,\dots  ,n$ are
the complex coordinates of the particles), we will study closed trajectories
of length $t$, i.e. $ z(t)=z(0)$ or open ones ($z(0)$ fixed, $z(t)$ left
unspecified, i.e. integrated over).

More precisely, if we consider some given bounded domain $S$ of area
$\cal{S}$ and define the occupation time $T_j$ as the time spent inside $S$
by the $j^{{\rm th}}$ particle, our goal is to compute the joint probability
distribution $P(T_1,T_2,\dots ,T_n)\;(\equiv P(\{T_j\}))$. Similarly, $A_j$
and $\theta_j$ being respectively 
the area enclosed by the trajectory of the $j^{{\rm th}}$
particle and its winding angle around O, we will be interested in the joint
laws $P(\{A_j\})$ and $P(\{\theta_j\})$.

On general grounds, we expect that the various properties of the chain will
be strongly influenced by the free Brownian motion of the center of mass
(c.o.m.) of the chain. But will they strictly satisfy the same laws ? With
the same scaling variables ? And what about the correlations among the
different variables ? Before answering those questions, we first recall some
standard results concerning a planar Brownian particle with a diffusion
constant $D$  [4-8].

Results i) and ii) concern open trajectories when $t\to\infty$ while iii)
concerns closed trajectories and is valid for all times $t$:
\begin{enumerate}  
\item[i)] Kallianpur-Robbins' law \cite{4} for the probability  distribution
 of the occupation time $T$ of a bounded domain of area ${\cal S}$:

\be\label{1}
 P\Lp T'=\frac{4\; \pi\; D\; T}{ {\cal S}\ln t}\Rp =\theta (T')\; e^{-T'}
\ee

\item[ii)] Spitzer's law \cite{5}
 for the angle $\theta$ wound around a given  point:

\be\label{2}
   P\Lp \theta'=\frac{2\theta }{\ln t} \Rp =\frac {1}{\pi}  \, 
   \frac {1}{1+  (\theta') ^2 } 
\ee
with the characteristic function:

\be\label{3}
\left\langle  e^{i\lambda \theta'} \right\rangle = e^{- \vert \lambda  \vert} 
\ee

\item[iii)] L{\'e}vy's law \cite{6} for the area $A$ enclosed by the closed
 trajectory of the particle:

\be\label{4}  
   P\Lp A'=\frac{A}{2\; D\;  t}\Rp =\frac{\pi}{2} \, \frac{1}{\cosh^2 (\pi A')}
\ee   

\be\label{5}
\left\langle  e^{i B A'} \right\rangle = 
\frac{\left(\frac{B}{2}\right)}{ \sinh \left(\frac{B}{2}\right)  }
\ee

\end{enumerate}

The distributions i) and iii) have moments of all orders in contrast with
ii) that has none.

Those laws were discovered more than 40 years ago and since that time, many
refinements have been made. For instance, in \cite{7}, the authors found the
asymptotic ($t\to\infty$) joint law of the small ($\theta_-$) and big
($\theta_+$) windings. $\theta_-$( resp. $\theta_+$) are the angles wound
around O and only counted when $r$ is smaller (resp. greater) than some fixed
$r_0$ ($r$ is the distance separating the particle from O). With the
rescaled angles $\theta'_{\pm}=\frac{2\theta_{\pm}}{\ln{t}}$, the
characteristic function writes \cite{7}:

\be\label{6}
\left\langle  e^{i( \lambda_{-} \theta'_{-}  +
 \lambda_{+} \theta'_{+} ) } \right\rangle = \frac{1}
 {\cosh (\lambda_{+}) + \frac{\vert\lambda_{-}\vert }{ \lambda_{+}}
 \sinh( \lambda_{+})}
\ee
($\lambda_+=\lambda_-=\lambda$ \ gives back Spitzer's law (\ref{2})).

Remark that (\ref{2}) and (\ref{6}) don't depend on the diffusion constant.
 This is quite different from the Brownian
motion on a bounded domain surrounding O. In that case, we have \cite{9}:

\be\label{7}
\left\langle  e^{i\lambda \theta} \right\rangle
 = e^{-c D  \vert \lambda  \vert t}
\ee
where $c$ is a constant depending on the geometry and the boundary
conditions. Here, $D$, appears as a multiplicator of
$|\lambda|$. We will use this remark at the end of the paper.

\section{The free Rouse chain}

Starting our study, we consider the following
set of coupled Langevin equations:

\bea
      \dot z_1 &=& k \, (z_2 -z_1) +\eta_1 \nonumber \\
  \dot z_l   &=& k \, (z_{l+1} + z_{l-1} -2z_l ) +\eta_l \ , \quad 
  2\le l \le n-1  \label{8} \\
  \dot z_n &=&   k \, (z_{n-1}-z_n) +\eta_n \nonumber  
\eea
where $k$ is the spring constant and
$\eta_m$ ($\equiv\eta_{mx}+i\eta_{my}$) a gaussian white noise:

\bea
	\langle \, \eta_m(t) \,	\rangle   &=& 0  \nonumber \\
   \langle \, \eta_m(t)\, \eta_{m'}(t') \, \rangle
 &=& 2 \, \delta_{mm'} \, \delta(t-t')     \label{9}
\eea
(This noise would correspond to a $D=1/2$ diffusion constant if
 particles were free).

For the chain c.o.m., we get
$\dot{z}_G=\frac{1}{n}\left(\sum_{i=1}^n\eta_i\right)\equiv\eta_G$ with
$\langle\eta_G(t)\eta_G(t')\rangle=\frac{2}{n}\delta(t-t')$. The c.o.m. 
motion is free with  
$D=1/(2n)$.

Introducing the complex $n$-vector
 $\eta $, eq.(\ref{8}) can be written in a matrix form:

\be\label{10}
        \dot z=- \;  k\; \gM\; z+\eta
\ee
where $\gM$ is the tridiagonal $(n\times n)$ matrix:

$$
\gM = \left(
\begin{array}{ccccc}
 \ 1 & -1  & \ 0  & \cdots & \ 0  \\
 -1  & \ 2  & -1  &  \cdots  & \ 0  \\
  \ 0  & -1  & \ 2  &  \cdots  & \ 0   \\
  \vdots & \vdots   & \vdots  & \ddots & \vdots  \\
 \ 0  & \ 0  & \ 0  & \cdots & \ 1  
\end{array}
\right)
$$
with eigenvalues:

\be\label{11}
   \oj \, = \, 2 \, \left( 1-\cos\frac{\pi (j-1)}{n}  \right)
 \   ,  \quad  1 \le j  \le n   
\ee
($\omega_1=0$;  $\det'\gM \equiv \prod_{j=2}^n \oj =n$)

With the matrix $ {\bf \omega} ={\rm diag}(\omega_i)$, we can write:
\bea
          {\bf \omega } &=& \gR^{-1}\; \gM\; \gR  \label{12}  \\ 
 z &=& \gR \ Z \label{13}   
\eea
where $\gR$ is an orthogonal matrix
and the components of $Z$ are the normal coordinates, that we will widely
use in the sequel. From ${\bf R}_{j1}=\frac{1}{\sqrt{n}}$, $j=1,\dots,n$, we
deduce that $Z_1\;(=\sum_{i=1}^n z_i / \sqrt{n})$ is essentially the c.o.m. 
coordinate. Remark also that
$\sum_{i=2}^n\omega_i|Z_i|^2=\sum_{i=2}^n|z_i-z_{i-1}|^2=\;^t\overline{z}{\bf
M} z$.

Let us call ${\cal P}( z, z^{(0)},t)$ the 
probability for the chain to go from configuration 
$z^{(0)}$ at $t=0$ to $z$ at time $t$.
${\cal P}$ satisfies a Fokker-Planck equation \cite{10}:

\be\label{14}
   \partial_t {\cal P}=\left( \tdz \, k\, \gM\, z +
   \tdzb\, k\, \gM\, \bar z +2 \, \tdzb \, \dz
\right) {\cal P}
\ee
where $\dz$ (resp. $\dzb$) is a $n$-vector of components $\dzi$
 (resp. $\dzib$) and $\tdz$ (resp. $\tdzb$) is the transpose of
 $\dz$ (resp. $\dzb$). The solution can be written in terms of
a path integral ($\ddz \, \ddzb \, = \prod_{i=1}^n
{\cal D}{z_i}  {\cal D}{\bar {z_i}}$):

\bea
  {\cal P}(z,\zo,t) &=& \det \, \left( \, e^{tk\gM } \, \right) 
  \, \int_{\zo}^{z}\ddz\ddzb 
\exp\left( -\frac{1}{2}\int_0^t \,  ^t(\dot {\bar z} +k\gM \bar z)(\dot z +
 k\gM z)
  \, \dd \tau \right)
 \label{15}       \\
&\equiv & F(z,\zo )\; .\; G_0(z,\zo,t) \nonumber  
\eea
with

\bea 
           F(z,\zo)    &=&     e^{-\frac{k}{2}
  \left( ^t\bar z \; \gM\; z   -  ^t\zob \; \gM \;\zo  \right)}= \nonumber \\
  &=& e^{- \frac{k}{2}  \sum_{i=2}^n  \left(
    \vert z_i -z_{i-1}    \vert ^2 - \vert z_i^{(0)} -z_{i-1}^{(0)}\vert ^2
\right)  }   = e^{-\frac{k}{2}  \sum_{i=2}^n \oi \left(  \vert Z_i \vert ^2  -
 \vert Z_i^{(0)}\vert ^2 \right)  } 
\nonumber	    \\ 
  G_0(z,\zo ,t)  &=&  \int_{\zo }^z \ddz\ddzb 
\exp\left( -\frac{1}{2}\int_0^t \left( ^t\dot {\bar z}\; \dot z \, +\, k^2
\;  ^t\bar z\;  \gM^2 z - 2k\; {\rm Tr} \gM\right) \dd\tau \right)= \nonumber\\
  &=&
\left\langle \; z\; \Lv \; e^{-tH_0}\; \Rv \; \zo\; 
 \right\rangle   \label{16}   \\
      H_0   &=& -2 \, \tdzb\; \dz +\frac{1}{2} \, k^2 \ ^t{\bar z}
      \,  \gM^2\, z -k \; {\rm Tr} \gM  \label{17} 
\eea

In fact, ${\cal P}$, eq.(\ref{15}),
can be easily deduced from the gaussian
distribution of $\eta$ (use  (\ref{10});
det($e^{tk \gM }$) is  the functional
Jacobian for the change of variable $\eta\to z$ \cite{11}).

$G_0(z,z^{(0)},t)$ is most conveniently written in terms of the normal
coordinates $Z_i$ and $Z_i^{(0)}$, clearly exhibiting the free motion of the
c.o.m. \cite{12}:

\bea
 &&G_0(z,\zo ,t) =  \frac{1}{2\pi t}e^{-\frac{1}{2t}
 \vert Z_1-Z_1^{(0)} \vert^2} \times \nonumber\\
&& \times\ \prod_{i=2}^n\left( \frac{s_i\; e^{k\oi t}}{2\pi }
 \exp \left\{ -\frac{1}{2}\left(\overline{Z}_ic_i Z_i +
 \overline{Z}_i^{(0)}c_i Z_i^{(0)}
 -  \overline{Z}_i^{(0)}s_i Z_i -  \overline{Z}_is_i Z_i^{(0)}  \right)\right\}
\right) \label{18} \\ 
&&             s_i   = \frac{k\oi }{\sinh (k\oi t)} \ ;\ 
           c_i   = k\oi \coth (k\oi t)   \nonumber
\eea

When $kt\gg1$, we get, for $G_0$, the limiting expression:

\bea
 G_0^{\infty }(z,\zo ,t)  &=&       \frac{1}{2\pi t}e^{-\frac{1}{2t}
 \vert Z_1-Z_1^{(0)} \vert^2}\prod_{i=2}^n\left( \frac{k\oi }{\pi }
 e^{-\frac{k\oi}{2}(\vert Z_i\vert^2 + \vert Z_i^{(0)}\vert^2 ) }
\right) \label{19} \\ 
                 &\equiv& {\cal G}_0  (z,\zo ,t)\; .\; g_0  (z,\zo )
		 \nonumber
\eea
where ${\cal G}_0$ is the c.o.m. propagator and $g_0$ can be simply written in
 terms of the $z_i$:

\be\label{20}
g_0  (z,\zo )= n\left( \frac{k}{\pi }   \right)^{n-1} e^{-\frac{k}{2}
   \sum_{i=2}^n  \left(
    \vert z_i -z_{i-1}    \vert ^2 + \vert z_i^{(0)} -z_{i-1}^{(0)}\vert ^2
\right)      }
\ee

 Furthermore, as can be
 easily checked, ${\cal P}$ is properly normalized:

\noindent$\int {\rm d}z{\rm d}{\bar z}\;{\cal P}(z,z^{(0)},t)=1$ 
($G_0$  given by (\ref{18}) or (\ref{19})).

 Now, we turn to the computation of the joint law $P(\{T_j\})$.

\section{Occupation times distribution}

\indent
Recall that $T_j$ is the time spent by particle $j$ inside a bounded domain
$S$ of area ${\cal S}$. We consider trajectories starting at $t=0$ from some
given configuration $z^{(0)}$ and reaching at time $t$ the final
configuration $z$. Leaving $z$ unspecified, we have, with positive
$p_i$ 's:

\bea
&&\left\langle  e^{-\sum_{i=1}^n p_i T_i}\right\rangle =
 \det \, \left( \, e^{tk\gM } \, \right) \times\nonumber\\
&& \quad   \times  \int \dd z\dd {\bar z} 
  \, \int_{\zo}^z\ddz\ddzb 
\exp\left( - \int_0^t\left(\frac{1}{2} \, 
^t(\dot {\bar z} +k\gM \bar z)(\dot z +
 k\gM z) + V_P(z)\right)
  \, \dd\tau \right)  \label{21} \\
&& \qquad \qquad =  \int \dd z\dd {\bar z}\;
 F( z,\zo  )\; G_P(z,\zo ,t) \label{22} 
\eea
with

\be
 G_P(z,\zo ,t)= \left\langle z\; \Lv \; e^{-t(H_0+V_P)}\; \Rv \; \zo
 \right\rangle  
\ee

\be\label{23}
   V_P(z)=\sum_{i=1}^n p_i {\bf 1}_S(z_i)
\ee
${\bf  1}_S(z_i)$ is the indicatrix function of the domain $S$.
Symbolically, we write:

\be\label{24}
  G_P=\sum_{m=0}^{\infty }(-1)^m G_0(V_PG_0)^m   
\ee
with

\bea
 &&  G_0(V_PG_0)^m = \dtm \int \left( 
 \prod_{j=1}^m \dd\overline{z}^{(j)}\dd z^{(j)} \right)
 G_0(z, z^{(m)},t-t_m)\times\nonumber\\ 
 &&\quad   \times \ V_P(z^{(m)})  
 G_0(z^{(m)}, z^{(m-1)},t_m-t_{m-1})V_P(z^{(m-1)}) \ldots
 V_P(z^{(1)})   G_0(z^{(1)}, z^{(0)},t_1)   \label{25}
\eea
($z^{(j)}$ is the chain configuration at time $t_j$; ${\rm
d}\overline{z}^{(j)}{\rm d}z^{(j)}=\prod_{i=1}^n {\rm
d}\overline{z}_i^{(j)}{\rm d}z^{(j)}_i$).

Let us compute the contribution $N_m(t)$ of this generic term to (\ref{22}).
Integrating over $z$, we have:

\bea
 N_m(t)   &=& (-1)^m \dtm\int \left( 
 \prod_{j=1}^m \dd\overline{z}^{(j)}\dd z^{(j)} \right)
 F(z^{(m)},z^{(0)}) \times\nonumber\\
 &&\ \times \ V_P(z^{(m)})  
 G_0(z^{(m)}, z^{(m-1)},t_m-t_{m-1}) \ldots
 V_P(z^{(1)})   G_0(z^{(1)}, z^{(0)},t_1) \label{26}\\
         &\equiv &  (-1)^m\int_0^t \dd t_m\int \left( 
 \prod_{j=1}^m \dd\overline{z}^{(j)}\dd z^{(j)} V_P(z^{(j)})    \right)
  F(z^{(m)},z^{(0)}) \Phi (t_m,\{ z^{(l)} \} ) \label{27}
\eea
$\Phi$ is a time convolution product of the free propagators $G_0$.
Disregarding for the moment the spatial integrations, the above expression is
well-suited, in the limit $t\to\infty$, for applying Tauberian theorems
\cite{13}. Introducing the Laplace Transform $\widehat{\Phi}$:

\be\label{28}
\widehat{\Phi }(u,\{ z^{(l)} \})\equiv
\int_0^{\infty } e^{-ut'} \Phi (t',\{ z^{(l)} \})\dd t'=
\prod_{k=1}^m \widehat{G_0}( z^{(k)}, z^{(k-1)}, u ) 
\ee
we notice that, when $u\to0^+$:

\be\label{29}
\widehat{G_0}( z^{(k)}, z^{(k-1)}, u )\equiv \int_0^{\infty } e^{-ut'}
G_0( z^{(k)}, z^{(k-1)}, t') \dd t' \sim
 \int_a^{\infty } e^{-ut'}
G_0( z^{(k)}, z^{(k-1)}, t') \dd t'
\ee
for some large $a$. So, we can use the asymptotic form $G_0^\infty$ in the
computation of $\widehat{G_0}$ and get:

\be\label{30}
\widehat{G_0}( z^{(k)}, z^{(k-1)}, u )\sim_{u\to 0^+}
\ln \Lp \frac{1}{u}\Rp \frac{1}{2\pi } g_0 \left(  z^{(k)}, z^{(k-1)}  \right)
\ee

A weak Tauberian theorem \cite{13} gives
 for the time integration in (\ref{27}):

\be\label{31}
\int_0^t \dd t_m  \Phi (t_m,\{ z^{(l)} \} ) \sim_{t\to \infty }
   \left(\frac{\ln t}{2\pi }\right)^m 
  \prod_{k=1}^m g_0 \left(  z^{(k)}, z^{(k-1)}  \right)
\ee

Finally, the result for $N_m(t)$ is:
\be\label{32}
N_m(t)  \sim_{t\to \infty } (-1)^m \left(\frac{n L_P \ln t }{2\pi }\right)^m 
\ee
\be\label{33}
L_P= \int \Lp \prod_{i=1}^n \dd z_i \dd\bar{z_i}\Rp V_P(z) 
\left(\frac{k}{\pi } \right)^{n-1}
e^{-k\sum_{i=2}^n \vert z_i -z_{i-1} \vert^2 } 
\ee
$L_P$ is computed with $V_P$, eq.(\ref{23}):
$L_P=\left(\sum_{i=1}^np_i\right){\cal S}$ 

Rescaling the occupation times $T_i$:
\be\label{34}
T'_i=\frac{2\; \pi\; T_i}{n \; {\cal S}\ln t}
\ee
we get:
\be\label{35}
\left\langle  e^{-\sum_{i=1}^n p_iT'_i}  \right\rangle =
\sum_{m=0}^{\infty }  (-1)^m \Lp \sum_{i=1}^n p_i \Rp ^m =
 \frac{1}{ 1+ \Lp  \sum_{i=1}^n p_i \Rp  }
\ee
This relationship is actually valid, by analytic continuation, for all the
positive $p_i$ 's (and not only when $\sum p_i <1$). This is because the
distribution $P(\{T_j\})$ has moments of all orders and consequently
$\langle e^{-\sum_{i=1}^n p_i T_i}\rangle$ is holomorphic when 
${\rm Re} (p_i)\ge 0$.

(\ref{35}) leads to the probability distribution:
\be\label{36}
P(\{ T'_i \} ) = \theta (T'_1)\; e^{-T'_1}\; \prod_{i=2}^n 
\delta( T'_i - T'_{i-1} )
\ee
($n=1$ gives back the Kallianpur-Robbins' law).

So, in the large time limit, the $T_i$ 's are strongly correlated leading to
identical $(T_i')$ 's. Moreover, we remark that
$T_i$ scales  like $n$ and, also, that the law
 for the c.o.m. would be the same as for one
monomer  (compare (\ref{34}) to -- (\ref{1}) with $D=1/(2n)$):
 the c.o.m. free motion dominates this process.

We also got similar exponential distributions for the rescaled variables
$T'$ in the following cases:
\begin{enumerate}
\item[i)] $T$ is the time spent when the {\it whole} 
 chain is inside $S$. $L_P$,
eq.(\ref{33}), is now computed with $V_P(z)=p\left(\prod_{i=1}^n{\bf 1}_S(z_i)\right)$. 
Introducing  
\be\label{37}
w({\cal S})=\int  \Lp \prod_{i=1}^n \dd z_i \dd\bar{z_i} {\bf 1}_S(z_i) \Rp  
e^{-k\sum_{i=2}^n \vert z_i -z_{i-1} \vert^2 } 
\ee
the rescaled variable writes:
\be\label{38}
T'= \Lp\frac{\pi }{k}\Rp ^{n-1} \frac{2\; \pi\; T}{n\; w({\cal S})\; \ln t }   
\ee
In contrast with (\ref{34}), $k$ is now present in the asymptotic law. For
instance, if $S$ is a small disk of radius $r_0$ ($k r_0^2 \ll1$), then
$w({\cal S})\sim{\cal S}^n$ and $T$ scales like $k^{n-1}$: when $k$ grows,
the chain collapses and it is easier to confine it inside a given domain.
\item[ii)]$T$ is the time spent by the chain when {\it at least} one of its  
particles is inside $S$. $T'$ is similar to (\ref{38}) except that
$w({\cal S})$ 
must be changed: $L_P$ is now computed with
$V_P(z)=p\left(1-\prod_{i=1}^n(1-{\bf 1}_S(z_i))\right)$.
\end{enumerate}

\vskip.5cm

To conclude this section, let us draw two lessons:
\begin{enumerate}
\item[i)] The scaling variables take the same general form as for the free
Brownian particle. In the sequel, we will show that it is still true for the
other quantities we study.
\item[ii)] For the computation of the perturbation theory, when
$t\to\infty$, we can systematically use the asymptotic form $G_0^\infty$ of
the unperturbed propagator. Obviously, for this consideration to hold, we
must be sure that the perturbation series is well behaved. In those
conditions, we will make a wide use of this remark.
\end{enumerate}

\section{Areas distribution}

\indent
Now, we compute the areas distribution $P(\{A_j\})$ for closed trajectories
of length $t$ starting and ending at some fixed $z^{(0)}$. To do so, we
insert the constraint:
\be\label{39}
\prod_{j=1}^n \delta \left(A_j -\frac{1}{4i}
\int_0^t(z_j  \dot{\bar z_j} -  \bar z_j \dot {z_j}) \dd\tau \right)
\ee
in the measure (\ref{15}) and use the relationship
$\delta(x)=\frac{1}{2\pi}\int e^{iBx}{\rm d}B$. It is easy to show that this
manipulation amounts to add $n$ different magnetic fields $B_j$ to the
initial system. Those fields are uniform, orthogonal to the motion plane and
such that particle $j$ is submitted to $B_j$.

With the $(n\times n)$ diagonal 
matrix $\bf B$ (${\bf B}_{ij}=B_i\delta_{ij}$), we get
\bea
 P(\{ A_i \}) &=&  \int
 \left(  \prod_{j=1}^n \frac{\dd B_j}{2\pi } e^{iB_jA_j}
   \right)
   \left( \frac{ G_{\bb }(\zo,\zo ,t)}{G_0(\zo ,\zo ,t)} \right) 
   \label{40}  \\
 {\rm with}\quad  
 G_{\bb }(\zo,\zo ,t)   &=& \left\langle \; \zo \; \Lv \;  e^{-tH_{\bb }} \;  
  \Rv\;   \zo \; \right\rangle 
  \label{41}  \\
       H_{\bb }     &=&  H_0 + V_{\bb }  \label{42}  \\
V_{\bb }(z) &=& \frac{1}{2}\left( - ^tz\, \bb\, \dz +  ^t\bar z\, \bb \, \dzb
    \right)   +\frac {1}{8} \,   ^t{\bar z} \, \bb^2 \, z    \label{43} \\
  &=& \frac{1}{2}\left( - ^tZ\, \bb'\, \partial_Z +  ^t{\bar Z}\, \bb' \,
   \partial_{\bar Z}
    \right)   +\frac {1}{8} \,   ^t{\bar Z} \, \bb'' \, Z  \label{44}   
\eea
(${\bf B'}={\bf R}^{-1}{\bf B}{\bf R},\;{\bf B''}={\bf R}^{-1}{\bf B}^2{\bf R}$).
In principle,  $P(\{A_j\})$ depends on $z^{(0)}$ but we will show that,
actually, this is not the case when $t\to\infty$ (remark that, for all $t$,
$P(\{A_j\})$ doesn't depend on the c.o.m. of $z^{(0)}$: this is due to
translation invariance and this is the reason why we consider the propagator
and not the partition function that would diverge like the area of the
plane, leading to serious problems in the perturbation theory).

Now, let us sketch the perturbative computation of $G_{\bf B}$. Following
our previous remarks, we will use $G_0^\infty$ for the unperturbed
propagator. The generic term writes:
\bea
&&(-1)^m \dtm \; \int \; \Lp  \prod_{j=1}^m
 \dd\overline{z}^{(j)} \dd z^{(j)} \Rp \ldots 
    \nonumber\\
&&\ldots G_0^{\infty } ( z^{(j+1)},  z^{(j)}, t_{j+1}-t_j  ) 
V_{\bb } ( z^{(j)})
G_0^{\infty } ( z^{(j)},  z^{(j-1)}, t_{j}-t_{j-1}  )\ldots  \label{45}
\eea

With normal coordinates, $V_{\bf B}(z^{(j)})$ takes the form:
\be\label{46}
  V_{\bb } ( z^{(j)})=\sum_{i,l=1}^n \Lp\frac{1}{2}\Lp
  - Z_i^{(j)}   \bb'_{il}   \partial_{ Z_l^{(j)}} + 
  \overline{Z}_i^{(j)}   \bb'_{il}        \partial_{\overline{Z}_l^{(j)}}
\Rp  +\frac{1}{8} \overline{Z}_i^{(j)}  \bb''_{il}          Z_l^{(j)} \Rp
\ee

We now proceed by inspection:
\begin{enumerate}
\item[i)] For the terms $Z_i^{(j)}{\bf B}'_{il}\partial_{Z_l^{(j)}}$, only
will survive, after integration, those contributions with $i=j$. The same
holds for ${\overline Z}_i^{(j)}{\bf B}'_{il}\partial_{{\overline Z}_l^{(j)}}$.
Moreover, the diagonal contributions exactly cancel except when $i=j=1$.
\item[ii)] We reach the same conclusion for the terms
$Z_i^{(j)}{\bf B}''_{il}Z_l^{(j)}$ (non diagonal terms vanish after
integration. Diagonal contributions, $i=j>1$, are subleading -- compared to
$i=j=1$ -- when $kt\gg1$: we recover the fact that the process is dominated
by the c.o.m. motion).
\end{enumerate} 

We are finally left with an effective perturbation $V_{\bf B}^{\rm eff}$:
\be\label{47}
  V_{\bb }^{{\rm eff}}= \frac{1}{2}\Lp
  - Z_1   \bb'_{11}   \partial_{ Z_1} + 
  \overline{Z}_1   \bb'_{11} \partial_{\overline{Z}_1}
\Rp  +\frac{1}{8}   \bb''_{11}        \vert  Z_1\vert^2 
\ee
Only the first mode is affected by the magnetic fields and we can disregard
the other modes that will cancel when taking the ratio $G_{\bf B}/G_0$.

Remark that 
\be\label{48}
\bp = \frac{1}{n}\Lp  \sum_{i=1}^n B_i  \Rp      \ {\rm and} \ 
\bpp = \frac{1}{n}\Lp  \sum_{i=1}^n B_i^2  \Rp   
\ee

The effective hamiltonian for the remaining mode writes:
\be\label{49}
 H_{\bb }^{{\rm eff}}= -2  \partial_{ Z_1} \partial_{\overline{Z}_1}
+   \frac{1}{2}\Lp
  - Z_1   \bb'_{11}   \partial_{ Z_1} + 
  \overline{Z}_1   \bb'_{11} \partial_{\overline{Z}_1}
 +    \frac{1}{4}   (\bb'_{11})^2        \vert  Z_1\vert^2 \Rp  +
 \frac{1}{8}  \Lp \bb''_{11} - (\bp)^2   \Rp  \vert  Z_1\vert^2 
\ee
It describes the behavior of a charged particle submitted to an uniform
magnetic field ${\bf B}'_{11}$ and an harmonic oscillator of frequency
$\omega=\frac{1}{2}\sqrt{{\bf B}''_{11}-({\bf B}'_{11})^2}$. Using known
results about this problem \cite{14}, we immediately get:
\bea
\frac{ G_{\bb }( z^{(0)}, z^{(0)},t ) }{ G_0( z^{(0)}, z^{(0)},t ) }
&=& \frac{ t \sqrt{\bpp }  }{ 2\sinh\Lp \frac{t}{2}  \sqrt{\bpp }  \Rp }
   \times\nonumber\\
&&\times\exp\Lp  -\frac{  \sqrt{\bpp }
\Lp \cosh  \Lp \frac{t}{2}  \sqrt{\bpp }\Rp  - \cosh \Lp  \frac{t}{2}\bp
\Rp \Rp \vert Z_1^{(0)}   \vert^2     }
{  2\sinh\Lp \frac{t}{2}  \sqrt{\bpp }  \Rp   }
\Rp \label{50}
\eea

However, for our computation to be consistent, we must consider this
expression in the large time limit. This is readily done in rescaling the
areas $A'_i=A_i/t$ and doing $t\to\infty$. The final expression for the
characteristic function of $P(\{A'_i\})$ is quite simple:
\be\label{51}
\La e^{i\sum_{j=1}^n B_jA'_j }  \Ra = \frac{\sqrt{\bpp } }
{2\sinh\Lp \frac{ \sqrt{\bpp } }{2}  \Rp } 
\ee
(${\bf B}''_{11}=\frac{1}{n}\sum_{i=1}^n B_i^2$ ; when $n=1$, (\ref{51})
gives back L\'evy's result, eq.(\ref{5})). 

Owing  to the form of
${\bf B}''_{11}$, $P(\{A'_i\})$ will only be a function of the variable

\noindent $\sqrt{\sum_{i=1}^n(A'_i)^2} \; (\equiv A')$, showing clearly that
the ($A'_i$) 's are correlated. 
Its determination is reduced to the computation of the following integral
\cite{13}:
\be\label{52}
P(\{ A'_i \} ) \equiv P(A') = \Lp \frac{2n}{\pi }  \Rp^{n/2}
\frac{1}{(A')^{n/2-1} }\int_0^{\infty }J_{n/2-1} (A'r)
\frac{r^{n/2+1}}{\sinh (r)} \dd r
\ee
where $J_\nu$ is a Bessel function. Closed form expressions can be given for
odd $n$ values. For instance:
\bea
n=3 \quad   P(A')     &=&   \frac{3\pi }{2A'} \frac{\tanh (\pi \sqrt{3}A')}
   {\cosh^2 (\pi \sqrt{3}A') }      \label{53}\\
n=5 \quad   P(A')     &=&    \frac{5 }{4A'^3} \frac
{ \tanh  (\pi \sqrt{5}A') -   (\pi \sqrt{5}A') ( 1- 
 3  \tanh^2  (\pi \sqrt{5}A') ) }
  {\cosh^2 (\pi \sqrt{5}A') }   \label{54}
\eea

Now, if we consider the distribution of the sum of the areas
$\AAA =\sum_{i=1}^n A_i$,
 it is obtained by setting $B_j=B,\;\forall j$. (\ref{51})
leads to ($\AAA '=\AAA /t$):
\be\label{55}
 \La e^{iB\AAA '} \Ra =  \frac{B}{2\sinh \Lp \frac{B}{2} \Rp}
\ee
With (\ref{5}), we see that the sum of areas has, asymptotically, exactly the
same distribution as the area enclosed by a single Brownian particle. In
fact, we can compute $\langle e^{iB\AAA '}\rangle$ for all $t$ values (and not
only when $t\to\infty$). This is because, that time, the matrix ${\bf B}$
(${\bf B}_{ij}=B\, \delta_{ij})$ commutes with ${\bf M}$. So, we are left with
an $\{$ harmonic oscillator + uniform magnetic field $\}$ 
 problem for each normal
coordinate (except for $Z_1$, that only feels a pure magnetic field). We get
the result \cite{14}:
\bea
 \La e^{iB\AAA '} \Ra &=& \frac{B}{2\sinh \Lp \frac{B}{2} \Rp }
  \prod_{i=2}^n \frac{F_i(B)}{F_i(0)}    \label{56}\\
 F_i(B)   &=&  \frac{\oi'}{2\pi\sinh (t\oi')}
   \exp\Lp -   \frac{\oi'}{2\pi\sinh (t\oi')} \Lp
 \cosh (t\oi') -\cosh(B/2)     \Rp
   \vert  Z_i^{(0)}  \vert^2   \Rp   \label{57}\\
 \oi'    &=& \sqrt{ \oi^2 + \Lp \frac{B}{2t}    \Rp^2    }     \label{58}
\eea
We recover (\ref{55}) in the limit $t\to\infty$ \ 
($\prod_{i=2}^nF_i(B)/F_i(0)\to 1$ when $t\to\infty$).

\vskip.8cm

To close this section, it is interesting to consider the asymptotic law for
the area $A'_j$ ($=A_j/t$) enclosed by a given monomer $j$. (\ref{51}) gives:
\be\label{59}
\La e^{iB_jA'_j} \Ra =  \frac{B_j}
{2\sqrt{n}  \sinh \Lp \frac{B_j}{2 \sqrt{n} } \Rp}
\ee
It follows that $A_j$ satisfies L\'evy's law (\ref{5}) and 
scales like $\frac{t}{\sqrt{n}}$. Remark that the area
swept by the chain c.o.m., G, should scale like $\frac{t}{n}$. On the other
hand, for the same gaussian noise, we would get $A_j\sim t$ if particle $j$
was free (i.e. $k=0$). The actual scaling of $A_j$ is intermediate: this
particle moves more freely than G but it is embedded in the chain, thus not
completely free!

Those considerations allow us to give a more precise sense to the statement:
``the process is dominated by the c.o.m. motion''. This one is true as long
as we look at occupation times. However, when we study finer quantities like
areas, this sentence must be corrected. Similar (even more dramatic)
deviations will occur when we look at winding angles. 

To end up with areas, let us remark that the case of open trajectories can
be treated exactly along the same lines as the one developed here, without
additionnal difficulties (in particular, (\ref{47}) still holds). We will
not address this problem in the present work.

\section{Winding angles distribution}

\indent
The last part of this paper will be devoted to the distribution
$P(\{\theta_j\})$ ($\theta_j$ is the angle wound around O by particle $j$
during a time $t$). We consider the same conditions as for Spitzer's law
($z^{(0)}$, initial configuration, fixed, with $z_j^{(0)}\neq0, \;\forall j$ ;
$z$, final configuration, unspecified ; $t\to\infty$).

We want to proceed as before and insert the constraint:
\be\label{60}
\prod_{j=1}^n \delta \left(\tj -\frac{1}{2i}
\int_0^t \left( \frac {z_j  \dot{\bar z_j} -  \bar z_j \dot {z_j}}  
  {z_j\bar z_j}    \right) \dd\tau \right)       
\ee
in the Wiener measure (\ref{15}). We are now faced with the problem of $n$
harmonically bound particles submitted to the magnetic fields of $n$
different point-like vortices located at the origin. The corresponding
hamiltonian is:
\bea
  H_{\lambda }  &=& H_0 + V_{\lambda }      \label{600}                  \\
    V_{\lambda }   &=& \sum_{i=1}^n\li \left( \frac{1}{z_i}\dzib -  
	\frac{1}{\bar z_i}\dzi \right) + 
	\sum_{i=1}^n \frac{\li^2}{2z_i\bar z_i} 
	\label{61}
\eea
and the distribution $P(\{\theta_i\})$  is given by:
\be\label{62}
 P(\{ \ti \}) = \int\left(  \prod_{j=1}^n \frac{\dd\lj }{2\pi }
 e^{i\lj\tj } \right) \int  \dd z \dd\bar{z}
 \; F(z,z^{(0)})\; G_{\lambda } (z,z^{(0)},t)
\ee
\be\label{63}
 G_{\lambda } (z,z^{(0)},t)= \left\langle \; z \; \Lv \;  
    e^{-t   H_{\lambda }}  \;  \Rv \; z^{(0)} \; \right\rangle
\ee

Studying the limit $t\to\infty$, we cannot develop directly as before a
perturbation theory with $V_\lambda$: this is because the last term in
$V_\lambda$ gives a divergent contribution \cite{9}. Due to this term,
all the eigenfunctions of $H_\lambda$ must vanish in O at least as
$\prod_{i=1}^n|z_i|^{|\lambda_i|}$ ($\equiv U(z)$). So, we redefine those
eigenfunctions \cite{9}:
\be\label{64}
       \Psi = U \widetilde{\Psi }
\ee
The new hamiltonian acting on $\widetilde{\Psi}$ is
\bea
  \widetilde{H_{\lambda }}  &=& H_0 +    \widetilde{  V_{\lambda }  } 
  \label{65} \\ 
 \widetilde{V_{\lambda }}(z)  &=&  \sum_{i=1}^n
 \left(  (\li - \vli ) \frac{1}{z_i}\dzib -(\li +\vli ) 
 \frac{1}{\bar z_i}\dzi
 \right) \label{66} 
\eea
with a propagator $\widetilde{G_\lambda}$
\be\label{67}
 \widetilde{ G_{\lambda } }(z,z^{(0)},t)= 
 \La\; z\; \Lv\;  e^{-t   \widetilde{H_{\lambda }}} \;  \Rv\; z^{(0)}\; \Ra = 
\frac{ U(z^{(0)})  }{ U(z) } \; G_{\lambda } (z,z^{(0)},t)
\ee
($\widetilde{G_0}=G_0$). That time, the perturbation theory is properly defined
and we can compute the characteristic function:
\be\label{68}
C(\{ \lambda_j   \} )\equiv \La  e^{i\sum_{j=1}^n  \lambda_j \tj    }   \Ra
  = \int \dd z\dd\bar{z} \Lp    \prod_{j=1}^n\frac{ \vert z_j \vert ^{\vlj } }
  {  \vert z_j^{(0)} \vert ^{\vlj }   }  \Rp F(z,z^{(0)}) \;   
 \widetilde{ G_{\lambda } }(z,z^{(0)},t)
\ee
with, symbolically,
\be\label{69}
 \widetilde{ G_{\lambda } } = \sum_{m=0}^{\infty } (-1)^m \; G_0^{\infty }\;
   (  \widetilde{ V_{\lambda } } \; G_0^{\infty })^m
\ee

Using integration by parts and also the relationship
$\partial_{z_i}\left(\frac{1}{\overline{z}_i}\right)=\pi \delta(z_i)$, we
first calculated $C(\{\lambda_i\})$ up to $4^{{\rm th}}$ order in
$\widetilde{V_\lambda}$, with the result:
\bea
 && C(\{ \lambda_j   \} )    \sim     e^{X/2} \; D(X)     \label{70}      \\
 && D(X)  \  =  \ 1+ \nonumber\\
 &&\quad  +  \Lp\frac{n+1}{2}  \Rp 
  \Lp  \frac{-X}{1!} +\frac{X^2}{2!}n 
  -\frac{X^3}{3!}\Lp \frac{3n^2-1}{2}\Rp +\frac{X^4}{4!}(3n^3-2n)-\ldots \Rp  
  \label{71}      \\
 &&    \quad    X    =  \Lp  \sum_{i=1}^n \vli   \Rp  \ln t       \label{72}    
\eea
The prefactor $e^{X/2}$ comes out from $U(z)$ in (\ref{68}) when integrated
over the final configuration: it will be present at all orders of the
computation. Morover, (\ref{70}) suggests that $C(\{\lambda_i\})$ is only a
function of $X$: this is actually the case, as will be shown in the sequel.

Let us consider the $m^{{\rm th}}$ order term in (\ref{68},\ref{69})
 and suppose that we
integrate, first, over $z,\; z^{(m)},\;z^{(m-1)},\dots ,z^{(k+1)}$. Following
the computation step by step, it is not difficult to convince oneself that
the integration over $z^{(k)}$ involves expressions of the form:
\be\label{73}
\int \dd\overline{z}^{(k)} \dd z^{(k)} \phi (z^{(k)},T) 
 \; \widetilde{V_{\lambda }}(z^{(k)}) 
 \; G_0^{\infty }(z^{(k)},z^{(k-1)},t_k-t_{k-1})
\ee
\be\label{74}
  {\rm where } \quad  \phi (z^{(k)},T) =
  e^{ -\frac{\vert  Z_1^{(k)}\vert^2 }{2T}  }
\; e^{  -\frac{1}{2} \sum_{i=2}^n k\oi \vert  Z_i^{(k)}\vert^2 }
\ee
and $T=t_l-t_k,\;k+1\le l \le m$. Let us call $J_k$ the result of (\ref{73}).
In the limit of long times, it reads:
\bea
&& J_k =-\Lp  \sum_{i=1}^n \vli  \Rp \ \times\nonumber\\
&\times&   
 \Lp -\frac{n+1}{2(t_k-t_{k-1})}  \phi (z^{(k-1)}, t_k-t_{k-1})  +
  \frac{1}{T+t_k-t_{k-1}}  \phi (z^{(k-1)},T+ t_k-t_{k-1}) \Rp \label{75}
\eea

The $m$ successive spatial integrations produce the factor
$\left(\sum|\lambda_i|\right)^m$ and, at the end, we are left with time
integrals of the form:
\begin{equation}
I_m(i_{m-1},\dots,i_0)(t)=
\dtm\frac{e^{-\sum_{i=1}^m\frac{\alpha_i}{t_i}-\sum_{i=1}^{m-1}\frac{\beta_i}{t_{i+1}-t_i}}}{(t_{i_{m-1}}-t_{m-1})\dots(t_{i_{1}}-t_{1})t_{i_0}}
\label{76}
\end{equation}
with \  $\alpha_i,\;\beta_i\;>0$ \ and 
\begin{equation}
i_{m-1}=i_{m-2}=\dots=i_k=m\;;\;i_{k-1}=i_{k-2}=
\dots=i_l=k\;;\;i_{l-1}=i_{l-2}=\dots=i_{j}=l\;;\;\dots
\label{77}
\end{equation}

We have proved, step by step, that;
\begin{equation}
I_m(i_{m-1},\dots,i_0)(t)
\sim_{t\to\infty}\frac{\left(\ln{t}\right)^m}{\prod_{l=0}^{m-1}(i_l-l)}
\label{78}
\end{equation}

Those considerations show that, actually, $C(\{\lambda_i\})$ is only a
function of $X$. So, we can write $D(X)=\sum_{m=0}^\infty a_m X^m$, with
$a_0=1$ (see eq.(\ref{71})).

Moreover, with the help of the above equation (\ref{78}), and also looking at the tree
structure exhibited in eq.(\ref{75}), the following recursion relation can
be shown:
\begin{eqnarray}
a_m&=&y\;\sum_{k=0}^{m-1}\frac{a_k}{(m-k)!}\\ 
y&=&-\frac{n+1}{2}
\end{eqnarray}
It allows to write a closed form formula for $D(X)$:
\begin{equation}
D(X)=\frac{1}{1-y(e^X-1)}=\frac{e^{-X/2}}{\cosh(X/2)+n\sinh(X/2)}
\label{81}
\end{equation}

With (\ref{70}) and, also, a rescaling of the angles
$\left(\theta'_i=\frac{2\theta_i} {\ln{t}}\right)$, we  get the desired
characteristic function:
\begin{eqnarray}
\La e^{i\sum_{j=1}^n \lambda_j\theta'_j}\Ra &=&
\frac{1}{\cosh(u)+n\sinh(u)} \label{82}\\
u&=&\sum_{i=1}^n\vli 
\end{eqnarray}
(with $n=1$, we recover eq.(\ref{3})).

We consider (\ref{82}) as the main result of this paper.

Finally, Fourier transformation shows that $P(\{\theta'_j\})$ is an
``infinite sum of products of Spitzer's laws'' (!) with highly correlated
variables:
\begin{eqnarray}
P(\{\theta'_j\})=\frac{2}{n+1}\sum_{k=0}^\infty \left\{
\left(\frac{n-1}{n+1}\right)^k\left(\prod_{j=1}^n \frac{1}{\pi
(2k+1)}\frac{1}{1+\left(\frac{\theta'_j}{2k+1}\right)^2}\right)\right\} 
\label{84}
\end{eqnarray}
All the moments of this distribution are infinite (unless they trivially
vanish).

\vskip.3cm

For a given particle $j$ of the chain, we have:
\begin{equation}
\La e^{i \lj \tj'} \Ra = \frac{1}{\cosh(\lj)+n\sinh(\vlj)}
\label{85}
\end{equation}
that leads to:

\be\label{850}
P(\theta'_j)=\frac{2}{n+1}\sum_{k=0}^\infty \left\{
\left(\frac{n-1}{n+1}\right)^k \frac{1}{\pi
(2k+1)}\frac{1}{1+\left(\frac{\theta'_j}{2k+1}\right)^2}\right\} 
\ee
The difference with Spitzer's law is due to the presence of $n$ in the
denominator of (\ref{85}).

To shed some light on this problem, let us go back to the joint law
(\ref{6}) of small and big windings for the chain c.o.m.. What could we
expect for the corresponding windings of particle $j$? With little effort, we
can say that:
\begin{enumerate}
\item[i)] The big windings will be roughly the same for both (when the chain
is far from O, particle $j$ follows the c.o.m. and winds around O in the
same way). So, we keep $\lambda_+$ unchanged in (\ref{6}).
\item[ii)] The  small windings will be quite different. This is because
particle $j$ is artificially maintained in the vicinity of O: despite its
higher mobility, {\it it spends the same time} as the c.o.m. in a given domain
surrounding O. As a consequence, its small windings law will be broadened.
Assuming that the remark following (\ref{7}) holds, we get this broadening by
changing $|\lambda_-|$ into $n|\lambda_-|$ in (\ref{6}) ($n$ is the ratio of
the diffusion constants; of course, we don't say at all that (\ref{7}) is
the law of small windings!).
\end{enumerate}

Thus, our guess for particle $j$ is:
\begin{equation}
\La \; e^{i(\lambda_+\theta'_{j+} +\lambda_-\theta'_{j-})}\; \Ra =
\frac{1}{\cosh(\lambda_+)+n\frac{|\lambda_-|}{\lambda_+}\sinh(\lambda_+)}
\label{86}
\end{equation}
Setting $\lambda_+=\lambda_-=\lambda_j$, we recover (\ref{85}). We are
aware that this argument is strictly heuristic and that (\ref{86}) remains
to be proved. Nevertheless, we think that it allows to explain correctly the
presence of $n$ in (\ref{85}).

\section{Conclusion}

\indent
Let us briefly summarize this work. We have computed explicitly the
asymptotic joint laws of the occupation times, areas and winding angles of a
chain of harmonically bound Brownian particles. 

\vskip.1cm

For all these properties, we have shown 
 that the scaling variables take the same {\it general} form as for the
standard Brownian motion.
However, a detailed study reveals important specific features that reflect
 a subtle interplay between 
  the free c.o.m. motion -- that strongly influences the
whole chain properties -- and the relative freedom of a given particle of the
chain. For occupation times distributions, it appears that the c.o.m.
satisfies the same law as a given monomer; now, for the areas, the scaling
becomes slightly different and, finally, for the winding angles, the law
itself is changed.
Remark also that correlations are systematically present.

\vskip.1cm

Moreover, we
observe that the scaling variables and the laws are very different from
those met in our study of the attached Rouse chain \cite{15} (in the latter
case, $\theta\sim t,\;A\sim\sqrt{t}$, and the winding angles are
uncorrelated). These differences are not so surprising since, in that case, we
had no translation invariance.

\vskip2cm

One of us (O.B.) acknowledges Dr. G. Oshanin for drawing his attention to
this problem. 

\vfill\eject

\vskip1cm
{\bf e-mail:}

benichou@lptl.jussieu.fr

desbois@ipno.in2p3.fr

\end{document}